\documentclass[pra,aps,amsmath,amssymb,preprint]{revtex4-1}
\usepackage{dcolumn}% Align table columns on decimal point
\usepackage{bm}% bold math
\usepackage{color}
\usepackage{graphicx}
\usepackage{latexsym}
\usepackage{amssymb}
\usepackage{amsfonts}
\usepackage{soul}
\usepackage{color}
\usepackage{amsmath}

\begin{document}

%\preprint{AIP/123-QED}

\title{Tunable room-temperature spin galvanic and spin Hall effects in van der Waals heterostructures}% Force line breaks with \\

\author{L. Antonio Ben\'{i}tez$^{1,2}$}
\thanks{These authors contributed equally to this work}
\email{an.benitez7@gmail.com}
\author{Williams Savero Torres$^{1}$}
\thanks{These authors contributed equally to this work}
\email{williams.savero@icn2.cat}
%\author{\textit{et al.}}
\author{Juan F. Sierra$^{1}$}
\author{Matias Timmermans$^{1}$}
\author{Jose H. Garcia$^1$}
\author{Stephan Roche$^{1,3}$}
\author{Marius V. Costache$^1$}
\author{Sergio O. Valenzuela$^{1,3}$}
\email{SOV@icrea.cat}
\affiliation{$^1$Catalan Institute of Nanoscience and Nanotechnology (ICN2), CSIC and The Barcelona Institute of Science and Technology (BIST), Campus UAB, Bellaterra, 08193 Barcelona, Spain}
\affiliation{$^2$Universitat Aut\`{o}noma de Barcelona, Bellaterra, 08193 Barcelona, Spain}
\affiliation{$^3$Instituci\'{o} Catalana de Recerca i Estudis Avan\c{c}ats (ICREA), 08010 Barcelona, Spain}

%This line break forced with \textbackslash\textbackslash}%

\date{\today}% It is always \today, today,

\begin{abstract}
\textbf{Spin-orbit coupling stands as a powerful tool to interconvert charge and spin currents and to manipulate the magnetization of magnetic materials through the spin torque phenomena \cite{SOV2015,manchon2015,Fert2016,SOV2017,manchon2019}. However, despite the diversity of existing bulk materials and the recent advent of interfacial and low-dimensional effects \cite{rojassanchez2013,manchon2015,Fert2016}, control of the interconvertion at room-temperature remains elusive. Here, we unequivocally demonstrate strongly enhanced room-temperature spin-to-charge (StC) conversion in graphene driven by the proximity of a semiconducting transition metal dichalcogenide (WS$_2$). By performing spin precession experiments in properly designed Hall bars, we separate the contributions of the spin Hall and the spin galvanic effects. Remarkably, their corresponding conversion efficiencies can be tailored by electrostatic gating in magnitude and sign, peaking nearby the charge neutrality point with a magnitude that is comparable to the largest efficiencies reported to date \cite{rojassanchez2013,rojassanchez2016}.
Such an unprecedented electric-field tunability provides a new building block for spin generation free from magnetic materials and for ultra-compact magnetic memory technologies.
}

\end{abstract}

%\pacs{74.25.Op, 74.72.-h, 74.25.Fy, 74.40.+k}% PACS, the Physics and Astronomy
\maketitle

%Two-dimensional (2D) materials assembled in van der Waals heterostructures are increasingly used to engineer systems with a desired collective functionality by precisely choosing the materials and the sequence in which they are stacked. Such heterostructures have led to the observation of exciting physical phenomena, such as topological currents and the Hofstadter butterfly effect or unconventional superconductivity, and the creation of structures with potential for practical applications, such as photovoltaic and light emitting devices or opto-valleytronic spin injectors. %The spectrum reconstruction in graphene encapsulated by hexagonal boron nitride allowed the demonstration of topological currents and the Hofstadter butterfly effect. The possibility of creating a two-dimensional superlattice by controlling the twisting angle between two sheets of graphene have led to the discovery of unconventional superconductivity. Optically active heterostructures such as light emitting devices can be created by combining 2D semiconductors and graphene.

While classical spintronics relies on the generation and manipulation of spin-polarized electrical currents using ferromagnetic materials, the emerging field of spin orbitronics is driven by the creation of pure spin currents in non-magnetic materials by means of the spin-orbit interaction (SOI) \cite{SOV2015,manchon2015,Fert2016,SOV2017}. In the spin Hall effect (SHE) and the inverse spin galvanic effect (ISGE), a charge current generates a transverse spin current and a non-equilibrium spin density, respectively, that can be utilized, for example, to manipulate the magnetization of ferromagnets in non-volatile memory technologies \cite{garello2018}. Various categories of materials are being intensively investigated for such a purpose, including metals,  oxides and topological insulators \cite{SOV2015,manchon2015,Fert2016,SOV2017,manchon2019,bibes2007}. Two-dimensional (2D) atomic crystals constitute a unique platform to engineer materials with novel functionalities in the limit of ultra-compact device architectures \cite{geim2013,novoselov2016}. Because they consist of atomically thin planes, their electrical, optical and spin properties can be enriched and tailored by proximity effects. In recent years, magnetic correlations and strong SOI have been successfully imprinted onto graphene and observed at room temperature \cite{wangz2015,vanwees2017,LAB2018}. When graphene is in direct proximity to a transition metal dichalcogenide (TMDC), it preserves its electronic properties while acquiring a complex spin texture with out-of-plane and winding in-plane components (Fig. 1a) that result in anisotropic spin dynamics \cite{gmitra2015,gmitra2016,LAB2018,cummings2017,ghiasi2017}.  Theoretical studies further suggest that the unique nature of this SOI can enhance the graphene StC conversion efficiency \cite{garcia2017,offidani2017}, while the coexistence of SHE and IGSE is allowed by new SOI terms such as the valley Zeeman coupling \cite{gmitra2015,gmitra2016}. The observation of the SHE was originally reported in graphene decorated with adatoms and in contact with WS$_2$ \cite{avsar2014} but follow-up studies suggest that the detected signal was not spin related \cite{vanwees2015,fuhrer2015,vanTuan2016}. More recently, the SHE was observed in heterostructures comprising multilayer graphene and MoS$_2$ \cite{casanova2019}. However, because MoS$_2$ is believed to be conductive and the experiment does not allow control of the carrier density of neither graphene nor MoS$_2$, it is not possible to fully discriminate between StC conversion by proximity effects and in the bulk of MoS$_2$ \cite{casanova2019}.% A more recent experiment using graphene-WS$_2$ reports the observation of the ISGE at room temperature and gate modulation is observed at 4.2 K \cite{ghiasi2019}.

In this work, we demonstrate large, gate-tunable StC conversion driven by both the SHE and ISGE in monolayer graphene-WS$_2$ heterostructures at room temperature. The SHE and ISGE are found to coexist in a narrow energy region nearby the charge neutrality point (CNP), enabling the control of both phenomena by tuning the carrier density $n$. The StC conversion mechanisms manifest in the presence of large spin transport anisotropy, as a consequence of the inherited spin textures of the graphene states. The observed SHE conversion efficiency as a function of $n$ and temperature is well reproduced by theoretical calculations of the spin Hall conductivity.
%As depicted in Fig. 1a, the graphene modified by the proximity of a TMDC preserves its linear Dirac band structure \cite{ggmitra2015,mitra2016}. The spins acquire a winding in-plane component, either clockwise or counterclockwise. The spin winding is expected to induce an in-plane non-equilibrium spin density with polarization perpendicular to an applied current (Fig. 1a), a phenomenon known as inverse spin galvanic effect (ISGE) \cite{SOV2015}. The ISGE is often accompanied by the spin Hall effect (SHE), in which an electrical current generates a transverse pure spin current \cite{SOV2015}. %Despite the rising interest, the ISGE by proximity effects have yet to be observed in experiments.
%The observation of the SHE was reported in graphene decorated with adatoms and in contact with WS$_2$ \cite{avsar2014} but follow-up reports suggest that the detected signal was not spin related \cite{vanwees2015,fuhrer2015}. Theoretical modelling of the experimental geometry further demonstrate the existence of multiple background contributions that could mimic the SHE \cite{vanTuan2016}. More recent experiments have shown spin-to-charge conversion in heterostructures comprising multilayer graphene and MoS$_2$ \cite{casanova2019}. However, because of the unknown conductivity of MoS$_2$, it is not possible to discriminate between spin-to-charge conversion by proximity effects or in the bulk of MoS$_2$ \cite{casanova2019,yang2016,dankert2017}.

The measurement scheme and optical image of a typical device are shown in Figs. 1b and 1c, respectively \cite{SOV2006,SOV2007,kimura2007} (see Methods for fabrication details
and Supplementary Fig. 1 for detailed device schematics). A carefully designed experimental protocol based on spin precession \cite{WST2017} allows us to isolate the proximity-induced StC conversion in the modified graphene (graphene-WS$_2$) from both spurious phenomena and competing StC conversion occurring in the bulk of the TMDC \cite{casanova2019}. The device consists of a patterned graphene Hall cross with a WS$_2$ flake along one of the arms and ferromagnetic injector/detector electrodes (F1, F2, F3) across the other. An electric field $E$ along the graphene-WS$_2$ arm (and associated current $I$, Fig. 1b) generates a spin current and spin accumulation due to the SHE (red arrows) and a non-equilibrium spin density conveyed by the ISGE (blue arrow, see Fig. 1a). The spins, carrying information on the SHE and the ISGE, diffuse in the graphene and are detected by measuring the non-local voltage $V_{\mathrm{nl}}^\mathrm{F}$ at F1. Alternatively, a spin current in graphene can be generated by applying $I$ in F1. In this case, the spin current and non-equilibrium spin density that reach the graphene(gr)-WS$_2$ are converted into a voltage $V_{\mathrm{nl}}^{\mathrm{gr-WS_2}}$ by the reciprocal effects: the inverse spin Hall effect (ISHE) and the spin galvanic effect (SGE). In the linear regime, for phenomena that are Onsager reciprocals such as the SHE and the ISHE, $V_{\mathrm{nl}}^\mathrm{F}=V_{\mathrm{nl}}^{\mathrm{gr-WS_2}}=V_{\mathrm{nl}}$.

For the StC conversion experiments, F1 and F2 are equivalent and either of them can be used to investigate the SHE, the ISGE and their reciprocals. In addition, F2 and F3 (in combination with F1) are used to independently characterize the spin dynamics in both the pristine graphene and the graphene-WS$_2$ \cite{LAB2018} (Supplementary Section I). This allows us to determine the StC conversion efficiencies without extra fitting parameters. The out-of-plane and in-plane spin relaxation lengths in graphene-WS$_2$, $\lambda_\mathrm{s}^\perp$ and $\lambda_\mathrm{s}^\parallel$, are about one-micrometre and few-hundred nanometres, respectively \cite{LAB2018}. Thus, the width $l_\mathrm{w}$ of the WS$_2$ flake is selected to be  $l_\mathrm{w}\sim1$ $\mu$m to simultaneously detect the SHE and the ISGE \cite{SOV2015}. Sufficiently large distances from the WS$_2$ to F1 and F2 ensure large-angle spin precession at low magnetic fields, for which the magnetization of the electrodes $\mathbf{M}_{1,2}$ remain unaffected \cite{BR2016}. A back-gate voltage $V_{\mathrm{g}}$ is applied to the substrate to tune the carrier density of both the graphene and WS$_2$.

For graphene-WS$_2$ in the $xy$-plane and $I$ applied along $\hat{\mathbf{y}}$ (Fig. 1b), the spins generated by the SHE are polarized in the $\hat{\mathbf{z}}$ direction while those generated by the ISGE are in the $\hat{\mathbf{x}}$ direction (see Fig. 1a). Because the ISGE and SHE spins are orthogonal, their contributions to $V_{\mathrm{nl}}$ can be readily separated using spin precession \cite{WST2017}. $V_{\mathrm{nl}}$ is proportional to the local spin accumulation projected along the detector magnetization $\mathbf{M}_1$. Due to magnetic shape anisotropy, $\mathbf{M}_{1}$ tends to be aligned with the F1 length, which is set along $\hat{\mathbf{y}}$, orthogonal to the spins generated by both the SHE and the ISGE (Fig. 1b). Thus, at zero magnetic field, $V_{\mathrm{nl}} = 0$ since neither the SHE nor the ISGE contribute to $V_{\mathrm{nl}}$. % Whenever a magnetic field $B$ is applied, spin precession can result in a measurable $V_{\mathrm{nl}}$. %For a magnetic field $B_{\perp}$ perpendicular to the graphene plane, the ISGE generated spins are no longer perpendicular to $\mathbf{M}_{1}$ as they reach F1, leading to $V_{\mathrm{nl}}\neq 0$ (Fig. 1c). Because the SHE spins are parallel to $B_{\perp}$, they do not precess and remain perpendicular to $\mathbf{M}_{1}$, which implies that $V_{\mathrm{nl}}$ is solely dependent on the ISGE.
For a low magnetic field $B_{z}$ perpendicular to the graphene plane, $V_{\mathrm{nl}}$ solely depends on the ISGE. While the SHE spins are parallel to $B_{z}$ and remain unaffected, the ISGE spins undergo precession. Because the ISGE spins are no longer perpendicular to $\mathbf{M}_{1}$ when they reach F1, $V_{\mathrm{nl}}\neq 0$ (Fig. 1d). In contrast, for an in-plane magnetic field $B_{x}$, the ISGE spins remain unaffected so that $V_{\mathrm{nl}}$ is solely dependent on the SHE (Fig. 1e). Similar arguments can be applied to the ISHE and the SGE.

%An unambiguous demonstration of any spin transport effect typically requires the observation of spin precession as described above. However, spin-to-charge conversion in metals has also been established by controlling the orientation of the injected spins by means of a large external magnetic field that rotates $\mathbf{M}_{1}$ \cite{SOV2006,SOV2007,kimura2007}. For graphene-WS$_2$, this technique is mostly suitable to investigate the (I)SGE where the magnetic field $B_x$ is in plane and magnetoresistive effects are minimized. When $\mathbf{M}_{1}$ rotates and develops a component along $\hat{\mathbf{x}}$, the (I)SGE can be detected. Note that an out of plane component of the stray field from F1 can lead to magnetoresistance effects and a step-like feature that could mimic the (I)SGE response. Even though our devices are designed to reduce such stray fields by placing WS$_2$ several micrometres away from F1, spin precession measurements are necessary to discard any possible artifacts.

For the transport to occur only in the modified graphene, it is crucial to verify the insulating character of WS$_2$. Otherwise, a current in WS$_2$ could generate both out-of-plane and in-plane spins, e.g. through the SHE in WS$_2$  \cite{casanova2019}, which can be mistaken with spins associated to the proximity effects.
Figure 2a shows the current $I_{\mathrm{ds}}$ versus $V_{\mathrm{g}}$ when a constant driving voltage $V_{\mathrm{ds}}$ is applied between graphene and WS$_2$. $I_{\mathrm{ds}}$ increases sharply at $V_{\mathrm{g}}^\mathrm{T} \sim 12$ V, marking the $V_{\mathrm{g}}$ at which WS$_2$ becomes conducting. Therefore, to avoid spin absorption (spin transfer) to (from) WS$_2$, all StC conversion measurements are acquired at $V_{\mathrm{g}}< V_{\mathrm{g}}^\mathrm{T}$.
Figure 2b shows the gate-dependent resistance $R$ of the graphene-WS$_2$. In this region, the graphene is $n$-doped with the CNP at $V_\mathrm{g}^{\mathrm{CNP}}=-10$ V. The fact that $V_\mathrm{g}^{\mathrm{CNP}} < V_{\mathrm{g}}^\mathrm{T}$ allows us to study the characteristic StC conversion for both electrons and holes.

%Having established the experimental protocol, and its region of validity in our device, we now focus on the spin-to-charge conversion measurements.
The insets of Figs. 2c and 2d show typical $R_{\mathrm{nl}}^{\downarrow,\uparrow}=V_{\mathrm{nl}}^{\downarrow,\uparrow}/I$ versus applied magnetic field with $\mathbf{M}_1$ antiparallel ($V_{\mathrm{nl}}^{\downarrow}$) and parallel ($V_{\mathrm{nl}}^{\uparrow}$) to $\mathbf{\hat{y}}$ (see Fig. 1b). The current $I$ is applied at F1 and $V_{\mathrm{nl}}^{\downarrow,\uparrow}$ measured in the graphene across the graphene-WS$_{2}$ arm. The orientation of $\mathbf{M}_1$ is prepared prior to the measurements by applying a magnetic field along $\mathbf{\hat{y}}$ with a magnitude exceeding the F1 coercive field. The inversion of the magnetic-field dependence when $\mathbf{M}_1$ reverses demonstrates the spin-related origin of the signal. The main panels of Figs. 2c and 2d show $\Delta R_{\mathrm{nl}}\equiv R_{\mathrm{nl}}^{\uparrow}-R_{\mathrm{nl}}^{\downarrow}$. By calculating the difference between $R_{\mathrm{nl}}^{\uparrow}$ and $R_{\mathrm{nl}}^{\downarrow}$, any non-spin related component in $V_{\mathrm{nl}}^{\downarrow,\uparrow}$ is eliminated.

In Fig. 2c, the magnetic field is applied in-plane in the ISHE configuration, as represented in Fig. 1e; therefore $\Delta R_{\mathrm{nl}}$ corresponds to the ISHE signal $R_{\mathrm{ISHE}}$. In Fig. 2d, the magnetic field is applied out-of-plane in the SGE configuration, as represented in Fig. 1d, and thus $\Delta R_{\mathrm{nl}}\equiv R_{\mathrm{SGE}}$. The experimental data in Figs. 2c and 2d exhibit an antisymmetric spin precession lineshape \cite{WST2017}. The spins injected by F1 are parallel to $\mathbf{\hat{y}}$ and, consequently, only when the magnetic field generates a $\hat{\mathbf{z}}$ ($\hat{\mathbf{x}}$) spin component, a nonzero $R_{\mathrm{ISHE}}$ ($R_{\mathrm{SGE}}$) is detected. At low magnetic fields $R_{\mathrm{ISHE}}$ and $R_{\mathrm{SGE}}$ are approximately linear; they reach maximum magnitude at about $\pm50$ mT and then decrease.
%Indeed, $R_{\mathrm{ISHE}}$ satisfies the relation $R_{\mathrm{ISHE}} \propto \mathbf{j}_\mathrm{s} \times \mathbf{\hat{s}}$ with $\mathbf{j}_\mathrm{s}$ the spin current and $\mathbf{\hat{s}}$ the unit vector parallel to the spin polarization vector. Only when the magnetic field generates a spin component perpendicular to the plane, a nonzero $R_{\mathrm{ISHE}}$ is detected. Similarly, by symmetry arguments, $R_{\mathrm{SGE}}$ is nonzero only when the magnetic field generates a spin polarization component perpendicular to the graphene-WS$_{2}$ arm (Fig. 1xx).
The extrema in $R_{\mathrm{ISHE}}$ and $R_{\mathrm{SGE}}$ indicate an aggregate spin precession angle of $\pi/2$ when reaching the Hall cross. The asymptotic decrease to zero at larger magnetic fields ($>50$ mT) is associated to spin dephasing.

The StC conversion efficiencies can be obtained from the spin precession lineshapes in Figs. 2c and 2d. To reduce the number of unknown parameters, the spin dynamics outside and inside the graphene-WS$_{2}$ region is independently characterized as follows. %using measurements as those in Figs. 2c and 2d. %In order to complete the device characterization, the spin transport in the pristine graphene is investigated.
Figure 2e shows typical non-local spin precession measurements $r_{\mathrm{nl}}=V_{\mathrm{nl}}^{\downarrow\uparrow,\uparrow\uparrow}/I$ in pristine graphene using electrodes F1 and F3. Measurements are performed for antiparallel ($\downarrow\uparrow$) and parallel ($\uparrow\uparrow$) configurations of $\mathbf{M}_{1,3}$ and magnetic fields $B_{z}$ and $B_{x}$ (see Supplementary Fig. 2 for $B_z$ measurements). %The current is driven between F1 and N1 and the voltage measured between F3 and N2 (Fig. 1c). The black (red) curve corresponds to initially parallel (antiparallel) configuration of the F1/F3 magnetizations, $\mathbf{M}_{1,3}$. The relative orientation of $\mathbf{M}_{1,3}$ is prepared prior to the measurements by properly sweeping the magnetic field along $\hat{y}$ (Fig. 1b).
By fitting the results to the solution of the Bloch equations, the spin transport properties of the pristine graphene are extracted %, that is, the spin lifetime $\tau_{s,gr}$, the spin-diffusion coefficient $D_{s,gr}$ and the spin relaxation length $\lambda_{s,gr}=\sqrt{\tau_{s,gr}D_{s,gr}}$,
(Supplementary Fig. 2). Figure 2f shows the spin precession response in graphene-WS$_2$ as a function of $B_{x}$ using electrodes F1 and F2. The results highlight the strongly anisotropic character of the spin transport, in agreement with prior reports \cite{LAB2018}. In particular, they show that as $B_{x}$ increases, the corresponding $r_{\mathrm{nl}}$ becomes much larger than its value at $B=0$, which demonstrates that that the spin lifetime of spins in the plane of graphene-WS$_2$, $\tau_\mathrm{s}^\parallel$, is much smaller than the lifetime of spins pointing out of plane, $\tau_\mathrm{s}^\perp$. Their magnitudes are calculated by modelling the spin diffusion in the graphene and the graphene-WS$_2$ with the parameters extracted from Fig. 2e, yielding $\tau_\mathrm{s}^\parallel = (6\pm1)$ ps and $\tau_\mathrm{s}^\perp = (52\pm10)$ ps. The anisotropy ratio, defined as $\zeta=\tau_\mathrm{s}^\perp/\tau_\mathrm{s}^\parallel$ is $\zeta=8\pm 3$ (Supplementary Section II). With the spin dynamics fully characterized, the spin and spin-current densities can be calculated, at any position, from the analytical solution of the Bloch diffusion equation (Supplementary Section II).

At this point, the spin precession lineshapes in Figs. 2c and 2d can be readily obtained from the spin current density $j^z_\mathrm{s}$  associated to the spins projected in $\hat{\mathbf{z}}$ (ISHE) and the in-plane component of the spin density in $\hat{\mathbf{x}}$ (SGE) in the graphene-WS$_2$ region (Fig. 1b). The only unknown parameters are the corresponding StC conversion efficiencies, which manifest as scaling factors (Supplementary Section II). In the ISHE case, the efficiency is quantified with the spin Hall angle $\theta_{\mathrm{SHE}}$, which measures the conversion from $j^z_\mathrm{s}$ to a charge current density $j^y_\mathrm{c}$ \cite{SOV2015}. In the case of the SGE, where a spin density is converted into $j^y_\mathrm{c}$, several figures of merit have been proposed. In our experiments, the spin density leading to $R_{\mathrm{SGE}}$ results directly from the spin current density $j^x_\mathrm{s}$. Therefore, it is possible to adopt an SGE equivalent to $\theta_{\mathrm{ISHE}}$, namely $\alpha_{\mathrm{SGE}} \equiv j^y_\mathrm{c}/j^x_\mathrm{s}$. Because the latter conversion is from a 2D spin current into a 2D charge current, $\alpha_{\mathrm{SGE}}$ is dimensionless. This contrasts with the commonly used inverse Edelstein effect length $\lambda_{\mathrm{IEE}}$, which quantifies the SGE conversion efficiency from a three-dimensional spin current into a 2D charge current and has the dimension of a length \cite{rojassanchez2013}.

The solid lines in Figs. 2c and 2d represent the calculated responses from which room-temperature $\theta_{\mathrm{SHE}} \approx0.3 $ \%, and $\alpha_{\mathrm{SGE}} \approx 0.1$ \% are estimated. The agreement between the model and experiment is excellent considering that only one adjustable (scaling) parameter is used. For comparison purposes, one can convert $\alpha_{\mathrm{SGE}}$ and $\theta_{\mathrm{SHE}}$ into $\lambda_{\mathrm{IEE}}$ through $\lambda^{\mathrm{SGE}}_{\mathrm{IEE}}=\alpha_{\mathrm{SGE}}~\lambda^{\parallel}_\mathrm{s}\approx0.42$ nm and $\lambda^{\mathrm{SHE}}_{\mathrm{IEE}}=\theta_{\mathrm{SHE}}~\lambda^{\perp}_\mathrm{s}\approx 3.75$ nm, where $\lambda^{\parallel}_\mathrm{s} = $ 420 nm and $\lambda^{\perp}_\mathrm{s} = 1.25$ $\mu$m are the in-plane and out-of-plane spin relaxation lengths in graphene-WS$_2$. Remarkably, these values compare very favourably with those estimated for heavy metals as $\theta_{\mathrm{SHE}} ~\lambda^\mathrm{M}_\mathrm{s}$ ($\lambda^\mathrm{M}_\mathrm{s}$ the metal spin diffusion length), yielding 0.2 nm for Pt, 0.3 nm for Ta and 0.43 for W \cite{lesne2016}. Furthermore, $\lambda^{\mathrm{SHE}}_{\mathrm{IEE}}$ for graphene-WS$_2$ is larger than the $\lambda_{\mathrm{IEE}}$ reported for any device at room temperature, being one order of magnitude larger than in Bi/Ag interfaces, $\lambda_{\mathrm{IEE}}= 0.2-0.33$ nm \cite{rojassanchez2013}, and almost twice $\lambda_{\mathrm{IEE}} = 2.1$ nm in the topological insulator $\alpha$-Sn \cite{rojassanchez2016}.% and a factor 6 larger than in LaAlO$_3$/SrTiO$_3$ interfaces at 7 K \cite{lesne2016}. %relaxation time?

The ISGE and SGE can also be investigated when $\mathbf{M}_{1}$ rotates and develops a component along $\hat{\mathbf{x}}$ with sufficiently large $B_x$ \cite{SOV2006,kimura2007}. As observed in Fig. 2e, the magnetization rotation is evident for $B_{x} > 0.2$ T, with $\mathbf{M}_{1}$  (and $\mathbf{M}_{3}$) becoming fully aligned with $B_{x}$ for $|B_{x}|>0.3$ T. Therefore, a broad step in $R_{\mathrm{nl}}$ is expected for $B_{x}$ between -0.3 and 0.3 T. Figure 3a shows $R_{\mathrm{nl}}$ versus $B_{x}$ for three representative gate voltages: at the CNP ($V_\mathrm{g} = -10$ V) and for electron ($V_\mathrm{g} = -3$ V) and hole ($V_\mathrm{g} = -13$ V) conduction. %The current is injected in F1 and $V_{\mathrm{nl}}$ measured in the graphene-WS$_2$ arm.
The SGE and the ISHE signals are clearly observed. The solid lines represent $M_{1x}$, the projected $\mathbf{M}_{1}$ along $\hat{\mathbf{x}}$, which is independently extracted using measurements as in Fig. 2e \cite{SOV2006}. The spin precession response associated to the ISHE, as shown in Fig. 2c, appears superimposed. The agreement between the field-dependence of $M_{1x}$ and the SGE step is excellent. In addition, the magnitude of the SGE extracted from these measurements coincides with the results obtained from spin precession (Fig. 2d); this allows us to discard any artifacts associated with stray magnetic fields. %Note that $R_{\mathrm{SGE}}$ is independent of the initial orientation of $\mathbf{M}_{1}$, while $R_{\mathrm{ISHE}}$ changes sign. This implies that the SGE and ISHE signals can be extracted from the sum and the difference of $V_{\mathrm{nl}}^{\downarrow}$ and $V_{\mathrm{nl}}^{\uparrow}$, respectively.

The results in Fig. 3a suggest that the ISHE and the SGE follow distinct carrier density dependences. While the SHE signal is clearly distinguished at the CNP, the SGE signal changes sign between electrons and holes and becomes undetectable at the CNP. Figure 3b represents the room-temperature magnitude of the ISHE and SGE versus $V_\mathrm{g}$. The SGE is quantified by means of the step height, $R^*_{\mathrm{SGE}}\equiv R_{\mathrm{nl}}(0.4 ~ \mathrm{T})-R_{\mathrm{nl}}(-0.4 ~ \mathrm{T})$ and the ISHE by the substraction $R^*_{\mathrm{ISHE}}$ between the values of $R_{\mathrm{ISHE}}$ at its two extrema (see Fig. 3a). It is observed that $n$ has a dramatic influence in the StC conversion efficiency. $R^*_{\mathrm{ISHE}}$ displays a sharp peak with its maximum located nearby the CNP; $R$ vs $V_\mathrm{g}$ is shown in Fig. 3b for comparison (solid black line). In contrast, $R^*_{\mathrm{SGE}}$ is asymmetric about the CNP, changing sign between electrons and holes, as concluded from Fig. 3a.

The change in sign in $R^*_{\mathrm{SGE}}$ stems from the change in the nature of the carriers, as the winding of the spin texture is symmetrical about the CNP \cite{gmitra2016,offidani2017}. This further confirms our interpretation of the signal as originating from the SGE. %On the other hand, the SHE arises from a combination of an intrinsic SOI and a valley Zeeman SOI, which is energy dependent and differs for electrons and holes.
Theoretical calculations for graphene-WS$_2$ suggest that $R_{\mathrm{ISHE}}$ also changes sign nearby the CNP with a behaviour qualitatively similar to that of $R_{\mathrm{SGE}}$. However, the magnitude of $R_{\mathrm{ISHE}}$ is expected to be much larger for holes than for electrons \cite{garcia2017}. As a consequence, the smearing effect resulting from the presence of charge puddles and room-temperature broadening might suppress the extremum in the electron side, leaving only the dominant hole peak by the CNP. The results in Fig. 4a, showing $R^*_{\mathrm{ISHE}}$ versus $n$ as the temperature $T$ decreases, indicate that this hypothesis is plausible. Indeed,  $R^*_{\mathrm{ISHE}}$ at room temperature (300 K) presents an incipient electron-hole asymmetry, with a steeper decrease for electron conduction. At 200 K, the decrease becomes steeper as thermal broadening is reduced. At 100 K a negative minimum at $n = 5\times 10^{11}$ cm$^{-2}$ becomes fully developed. Figure 4b shows the computed spin Hall conductivity $\sigma^{\mathrm{SHE}}_{xy}$ in the weak disorder limit \cite{garcia2017} (see Supplementary Section III). The parameters used for the calculations to match the experimental trends are within a factor two of those obtained by density functional theory \cite{gmitra2016,garcia2017}. Notably, the theoretical model describes qualitatively all of the experimentally observed features, including the relative magnitude of the positive and negative extrema, their approximate width and the temperature $T \sim 200$ K at which the change in sign is observed. Although the temperature dependence of $R^*_{\mathrm{ISHE}}$ is highly non-trivial, $R^*_{\mathrm{ISHE}}$ varies roughly as $1/T$ nearby the extrema (Inset Fig. 4a), which is also well reproduced by the calculations (Inset Fig. 4b). The tendency for saturation in $R^*_{\mathrm{ISHE}}$ at low $T$ for hole conduction, which is not observed in the model, could indicate the presence of charge puddles or other types of impurities and defects.

The observation of the SGE and ISHE within the gap of WS$_2$, their dependence on $n$ and the excellent agreement with theoretical expectations prove that the phenomena originate from SOI proximity effects in graphene. However, the magnitude of the SGE and ISHE and the large spin relaxation anisotropy suggest that the theoretical understanding is still incomplete and deserves further investigation. Theoretical results, which consider only intrinsic effects \cite{garcia2017,offidani2017}, are compatible with the experimental $\theta_{\mathrm{SHE}}$ only for weak intervalley scattering while yielding an $\alpha_{\mathrm{SGE}}$ that is at least one order of magnitude smaller than the experimental value. Furthermore, a large SHE, as observed in Fig. 2c, has been deemed incompatible with anisotropic spin relaxation (Fig. 2f). Anisotropic spin relaxation signals the presence of intervalley scattering, which effectively reduces the mass of the carriers and leads to the suppression of the SHE \cite{cummings2017,garcia2017}. Such discrepancies between experimental results and purely intrinsic effects suggest additional extrinsic contributions in the observed SHE and ISGE. A larger localized SOI can be mediated by localized defects, such as sulphur vacancies in WS$_2$, while an effective enhancement of $\theta_{\mathrm{SHE}}$ could arise from resonant scattering as proposed for metallic adatoms \cite{vanTuan2016,ferreira2014}.

Our experiments therefore provide valuable insights into the physics underpinning proximity effects in graphene. In addition, they offer novel strategies to manipulate spin information in ultra-compact van der Waals heterostructures, which could further impact the design and performance of advanced magnetic memory technologies \cite{garello2018}. Indeed, in recent years the SHE and ISGE have evolved from subtle academic phenomena to effective approaches to electrically manipulate the magnetization of a ferromagnet \cite{SOV2015,garello2018}. The electric-field control of the StC conversion, both in magnitude and sign, and the unprecedented efficiency thus open the door to magnetization control with 2D materials and to novel spin-logic circuits without the use of ferromagnets.

\textit{Note}: After completion of the current research, a work studying the spin galvanic effect using magnetization rotation in graphene-WS$_2$ has been reported \cite{ghiasi2019}. The investigation presents results showing SGE modulation at 4.2 K and large $n$ and shows that the SGE signal persists up to room temperature. However, the SGE modulation and the SHE are not observed at room temperature.

.

\section{Methods}
\small{\textbf{Device Fabrication.}
van der Waals graphene-WS$_2$ heterostructures were fabricated by dry viscoelastic stamping \cite{LAB2018}. The transfer set-up comprises an optical microscope with large working distance optical objectives (Nikon Eclipse Eclipse LV 100ND) and a three-axis micrometer stage. Graphene is obtained by mechanical exfoliating highly-oriented pyrolytic graphite (SPI Supplies) onto a $p$-doped Si/SiO$_2$ substrate. Large-area monolayer graphene is selected by optical contrast after a previous calibration with Raman spectroscopy. To fabricate the van der Waals heterostructure, WS$_2$ flakes are transferred onto a viscoelastic stamp (Gelpack), which is then transferred on top of the graphene target. After assembling, the stacks are annealed for 1 hour at 500 $^{\circ}$C in high vacuum (10$^{-8}$ Torr). The heterostructure is coated with a PMMA resin mask and patterned into a Hall cross bar using electron-beam lithography followed by oxygen plasma etching \cite{ZMG2019}. The resin is then removed with acetone. The contact electrodes are defined in two electron-beam lithography steps, one for the normal metal electrodes, Ti(1nm)-Pd(50nm) and the other for the ferromagnets, TiO$_\mathrm x$/Co(30nm). The contact materials are deposited by electron-beam evaporation in a chamber with base pressure of 10$^{-8}$ Torr. The TiO$_\mathrm{x}$ barriers are fabricated by evaporating 4 {\AA} + 4 {\AA} of Ti and 30 min oxidation after each evaporation in an oxygen atmosphere of about 10$^{-1}$ Torr.

\textbf{Electrical characterization.}
The devices are wired to a chip carrier that is placed in a cryogen-free cryostat. Charge transport properties were characterized by means of two- and four-terminal measurements. The contact resistance in the TiO$_\mathrm x$/Co electrodes are larger than 10 k$\Omega$. The typical average electron/hole mobility is in the range of $\mu=5000$ cm$^2$V$^{-1}$s$^{-1}$ with a residual carrier density of $2.5\times 10^{11}$ cm$^{-2}$. A back-gate voltage applied to the $p$-doped Si subtrate is used to control the carrier density $n$ in the device.}

%\textbf{Analytical calculation of the spin Hall effect.}
%The proximity with a WS$_2$ monolayer modifies the electronic properties of graphene due to a combination of the periodic potential, originating from the incommensurability of the two lattices, and a  weak hybridization between the carbon and the tungsten atoms. This combination opens a small electronic gap and imprints in graphene an effective spin-orbit coupling, both effects can be captured by the following Hamiltonian \cite{gmitra2016} :
%\begin{equation}
%H_{\rm gr} =H_{\rm gr}+\Delta \sigma_z+ (\lambda_{\rm I}  s_z + \lambda_{\rm VZ}s_0 )\tau_z \sigma_z  + \lambda_{\rm R} ( \tau_z\sigma_x s_y- \sigma_y s_x ), \label{hamiltonian_model}
%\end{equation}
%where the first term is the Hamiltonian for pristine graphene, the second term opens an energy gap in the spectrum of magnitude $\Delta$, the third term is composed by the intrinsic and a valley-Zeeman spin-orbit coupling of strengths $\lambda_{\rm I}$ and $\lambda_{\rm VZ}$, respectively, and the last term is the Rashba spin-orbit coupling with magnitude $\lambda_{\rm R}$.

\textbf{Data Availability.} The data that support the plots within this paper and other findings of this study are available from the corresponding authors upon reasonable request.

\vspace{5mm}

\noindent \textbf{Acknowledgments} We thank D. Torres for help in designing Fig. 1. This research was partially supported by the European Union's Horizon 2020 research and innovation programme Graphene Flagship CORE 2, under grant agreement No. 785219, by the European Research Council under Grant Agreement No. 306652 SPINBOUND, by the Spanish Ministry of Economy and Competitiveness, MINECO (under Contracts FIS2015-62641-ERC, MAT2016-75952-R, and SEV-2017-0706 Severo Ochoa), and by the CERCA Programme and the Secretariat for Universities and Research, Knowledge Department of the Generalitat de Catalunya 2017 SGR 827. M.T. acknowledges support from the European Union’s Horizon 2020 research and innovation programme under the Marie Sklodoswa-Curie grant agreement Nº 665919.

\vspace{5mm}

\noindent \textbf{Author contributions}
L.A.B. and S.O.V. designed the device and L.A.B. and J.F.S. fabricated it. The measurements were performed by L.A.B. and W.S.T. with the participation of J.F.S. The experimental set-up was implemented by M.T. and M.V.C., who also helped with the measurements. J.H.G. and S.R. carried out the SHE quantum simulations. L.A.B., W.S.T. and S.O.V analyzed the data and wrote the manuscript. All authors contributed to the study, discussed the results and commented on the manuscript. S.O.V. supervised the work.

%S.O.V planned and supervised the work.
%The devices were fabricated by L.A.B. and J.F.S with the assistance of W.S.T. . Measurements were performed by L.A.B. and W.S.T. with the help of J.F.S., M.T. and M.V.C. The measurement set-up was prepared by M.T. and M.V.C. The quantum simulations in were carried out by J.H.G. and S.R. The manuscript was written by L.A.B., W.S.T. and S.O.V. All authors discussed the results and commented on the manuscript. S.O.V supervised the work.
%
%\vspace{5mm}
%
%\noindent \textbf{Additional Information} The authors declare no competing financial interests. Reprints and permissions information is available online at http://npg.nature.com/reprintsandpermissions. Correspondence and request for materials should be addressed to S.O.V. (SOV@icrea.cat).
%%%
\newpage
\begin{figure}[ht]
\vspace{5mm}\includegraphics[width=1\linewidth]{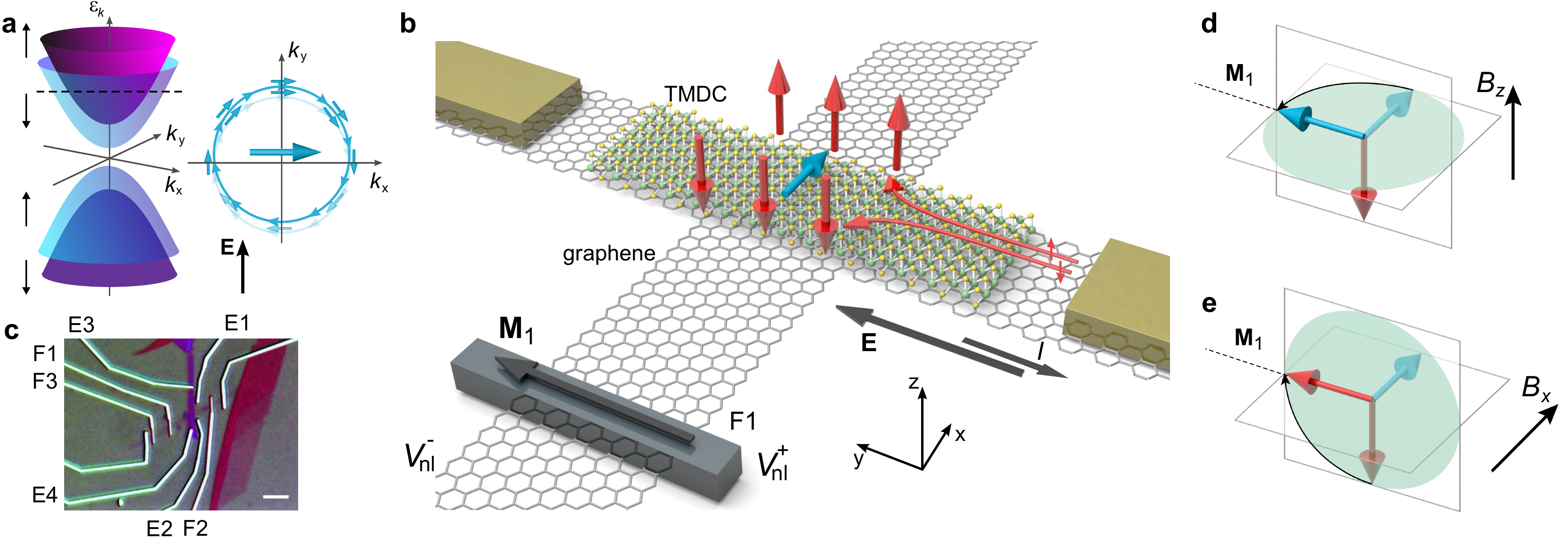}
\vspace{-10mm}
\caption{\textbf{Spin-to-charge conversion in graphene by proximity of a TMDC and measurement scheme}. \textbf{a}, Left: Spin-split bands with opposite spin helicity for graphene-TMDC (blue and purple). The black arrows represent the out-of-plane spin component. Right: Illustration of a microscopic picture of the ISGE. Spin texture for one of the bands developing a non-equilibrium spin density (blue arrow) in an applied electric field $E$. The opposite helical spin texture with lower Fermi wave vector, and smaller contribution to the spin density, is not drawn for clarity. \textbf{b}, Schematics illustrating the basic concepts of the measurement approach. The main elements of the device include a graphene Hall cross with a TMDC strip over one of the arms and a ferromagnet (F1) contacting the other. A current $I$ generated by an electric field $E$ in $\mathbf{\hat{y}}$, along graphene-TMDC, induces a non-equilibrium spin density due to the ISGE with spins along $\mathbf{\hat{x}}$ (blue arrow, see  \textbf{a}). Spin accumulation with spins out of plane (along $\mathbf{\hat{z}}$) are generated by the SHE with opposite orientation at opposite edges of the graphene-TMDC (red arrows). The induced spins diffuse in graphene towards F1 and are detected by measuring $V^\mathrm{F}_{\mathrm{nl}} = V_{\mathrm{nl}}^+- V_{\mathrm{nl}}^-$. At zero magnetic field $V^\mathrm{F}_{\mathrm{nl}}=0$, as the ISGE and SHE spins are perpendicular to the F1 magnetization $\mathbf{M}_1$. \textbf{c}, Optical image of a device with TMDC = WS$_2$. Besides F1 and the Hall cross, the device comprises additional ferromagnetic electrodes (F2, F3) and contacts to graphene (E1, E2) and the WS$_2$ (E3, E4).
%The additional contacts are used to characterize the spin transport in graphene (F3 and F1), the spin relaxation anisotropy in graphene-WS$_2$ and the conductivity of graphene-WS$_2$ and of WS$_2$ (E1-E4).
%A detailed schematics of the device is shown in Supplementary Fig. 1.
\textbf{d}, ISGE detection. A magnetic field $B_z$ induces spin precession on the ISGE generated spins (blue arrow), leading to a component along $\mathbf{M}_1$ and $V_{\mathrm{nl}}\neq0$. The SHE spins (red arrow) do not contribute to $V_{\mathrm{nl}}$ as they are parallel to $B_z$ and do not precess. \textbf{e}, SHE detection. Similar to \textbf{d}, a magnetic field $B_x$ induces spin precession on the SHE generated spins leading to $V_{\mathrm{nl}}\neq0$. The ISGE spins do not precess and do not contribute to $V_{\mathrm{nl}}$.}
\label{Fig1}
\end{figure}
\newpage
\begin{figure*}[ht!]
\vspace{-10mm}\hspace{15mm}\includegraphics[width=0.9\linewidth]{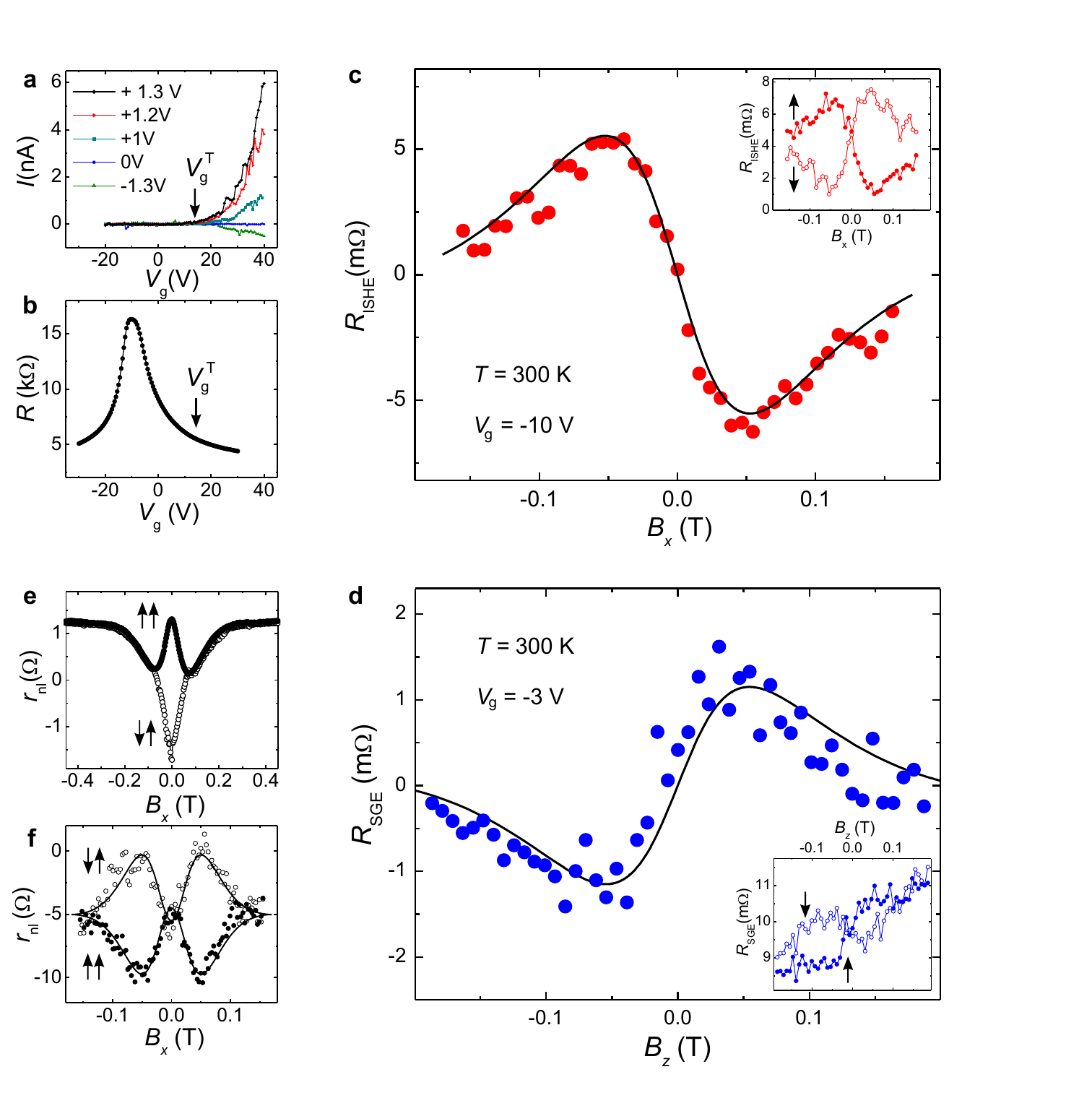}
\vspace{-8mm}
\caption{\textbf{Sample characterization and spin-to-charge conversion measurements}. \textbf{a}, Transfer characteristics $I_{\mathrm{ds}}$ versus $V_\mathrm{g}$ for different bias voltage $V_{\mathrm{ds}}$ in graphene-WS$_2$; $V_\mathrm{g}^\mathrm{T}$ indicates the back-gate voltage at which $I_{\mathrm{ds}}$ is first observed. \textbf{b}, Resistance $R$ vs. $V_\mathrm{g}$ along graphene-WS$_2$. The charge neutrality point is located at $V_\mathrm{g}\approx -10$ V, well below $V_\mathrm{g}^\mathrm{T}$. \textbf{c}, Representative room-temperature non-local measurements in the ISHE configuration. Inset: $R_{\mathrm{nl}}^{\downarrow,\uparrow}$ vs. $B_x$ for $\mathbf{M}_1$ antiparallel ($\downarrow$, open symbols) and parallel ($\uparrow$, solid symbols) to $\mathbf{\hat{y}}$. Main panel: $R_{\mathrm{ISHE}}=(R_{\mathrm{nl}}^{\uparrow} - R_{\mathrm{nl}}^{\downarrow})$. \textbf{d}, As in \textbf{c} in the SGE measurement configuration. \textbf{e}, Spin injection and detection using F1 and F3 vs $B_x$ (see Fig. 1c and Supplementary Fig. 1) for antiparallel (open symbols) and parallel (solid symbols) configuration of the electrodes magnetizations. \textbf{f}, As in \textbf{e} using F1 and F2. The signal becomes larger when $B_x\neq 0$ than at $B_x=0$, indicating a large spin relaxation anisotropy in graphene-WS$_2$ with long spin lifetimes out of the graphene plane. Solid lines in \textbf{c}, \textbf{d} and \textbf{f} are fittings to the solution of the anisotropic spin diffusion equations. Measurements in \textbf{a}, \textbf{b}, \textbf{e} and \textbf{f} are performed at 200 K.} \label{Fig2}
\end{figure*}
\newpage
\begin{figure*}[ht!]
\hspace{5mm}\includegraphics[width=1.05\linewidth]{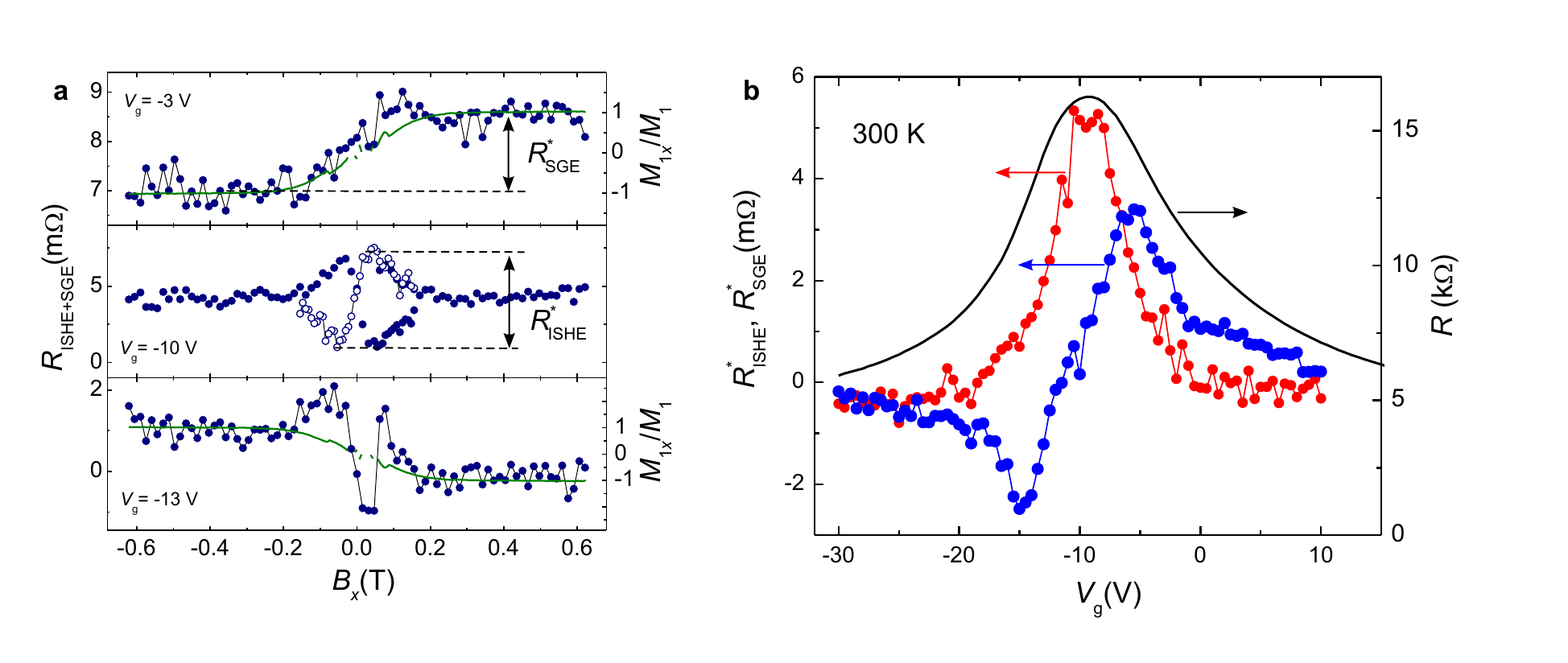}
\caption{\textbf{Gate control of the inverse spin Hall and spin galvanic effects at room temperature}. \textbf{a}, Nonlocal resistance $R_{\mathrm{nl}}$ vs $B_x$ at selected $V_\mathrm{g}$ showing the ISHE precession response at low magnetic fields and the step feature associated to the SGE. The latter is observed when the magnetic field is large enough to rotate $\mathbf{M}_1$ and generate a component along $\mathbf{\hat{x}}$, $M_1^x$. The normalized  $M_1^x$ (sine of the rotation angle) is extracted from spin precession measurements as in Fig. 2e \cite{SOV2006} and represented with green lines. \textbf{b}, Spin-to-charge conversion for the ISHE (red) and the SGE (blue) as a function of $V_\mathrm{g}$. The corresponding conversion efficiencies are quantified with $R^*_{\mathrm{ISHE}}$ and $R^*_{\mathrm{SGE}}$ as defined in \textbf{a}. $R$ vs. $V_\mathrm{g}$ is shown for comparison (black line).
} \label{Fig3}
\end{figure*}
\newpage
\begin{figure*}[ht!]
\includegraphics[width=0.7\linewidth]{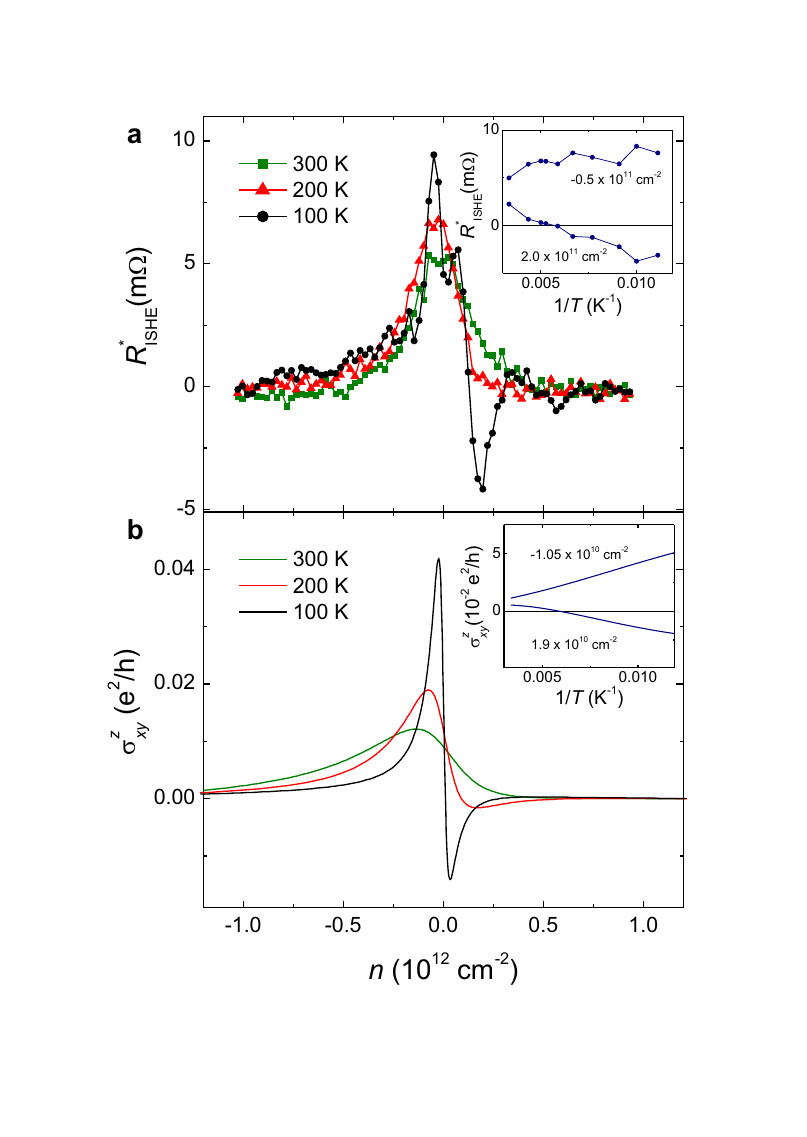}
\vspace{-20mm}
\caption{\textbf{Temperature dependence of the SHE spin-to-charge conversion}. \textbf{a}, $R^*_{\mathrm{ISHE}}$ vs. carrier density $n$ at the specified temperatures. \textbf{b}, spin Hall conductivity $\sigma_{xy}^z$ in the weak disorder limit using realistic parameters. The insets in \textbf{a} and \textbf{b} show the temperature dependence of $R^*_{\mathrm{ISHE}}$ (\textbf{a}) and $\sigma_{xy}^z$ (\textbf{b}) for $n$ at the extrema positions.
} \label{Fig4}
\end{figure*}
\end{document}